\documentclass[10pt]{article}

\usepackage{setspace} 
\usepackage{amsmath}
\usepackage[dvips]{graphicx}
\usepackage{amssymb}
\usepackage{color}
\usepackage{textcomp}
\usepackage{cite}
\usepackage{upgreek}
\usepackage{bm}

\usepackage[labelfont=bf,labelsep=period,justification=raggedright]{caption}

\makeatletter
\renewcommand{\@biblabel}[1]{\quad#1.}
\makeatother

\date{}

\begin{document}
	\begin{flushleft}
	{\Large
		\textbf{Influences of Excluded Volume of Molecules on Signaling Processes on the Biomembrane}
	}
	\\
	Masashi Fujii$^{\ast}$, 
	Hiraku Nishimori,
	Akinori Awazu
	\\
	{\bf} Department of Mathematical and Life Sciences, Hiroshima University, Higashi-Hiroshima, Hiroshima, Japan.
	\\
	$\ast$ E-mail: mfujii0123@hiroshima-u.ac.jp
	\end{flushleft}
	\section*{Abstract}
		
		We \textcolor{black}{investigate} the influences of the excluded volume of molecules on biochemical reaction processes on 2-dimensional surfaces using a model of signal transduction processes on biomembranes.
		We perform simulations of the 2-dimensional cell-based model, which describes the reactions and diffusion of the receptors, signaling proteins, target proteins, and crowders on the cell membrane.
		The signaling proteins are activated by receptors, and these activated signaling proteins activate target proteins that bind autonomously from the cytoplasm to the membrane, and unbind from the membrane if activated.
		If the target proteins bind frequently, the volume fraction of molecules on the membrane becomes so large that the excluded volume of the molecules for the reaction and diffusion dynamics cannot be negligible.
		We find that such excluded volume effects of the molecules induce non-trivial variations of the signal flow, defined as the activation frequency of target proteins, as follows.
		With an increase in the binding rate of target proteins, the signal flow varies by $i)$ monotonically increasing; $ii)$ increasing then decreasing in a bell-shaped curve; or $iii)$ increasing, decreasing, then increasing in an S-shaped curve.
		We further demonstrate that the excluded volume of molecules influences the hierarchical molecular distributions throughout the reaction processes.
		In particular, when the system exhibits a large signal flow, the signaling proteins tend to surround the receptors to form receptor-signaling protein clusters, and the target proteins tend to become distributed around such clusters.
		To explain these phenomena, we analyze the stochastic model of the local motions of molecules around the receptor.
		
	\section*{Introduction}
	
		Several living systems can sense and respond to environmental variations by means of internal biochemical processes.
		The adaptation of cells and the cell fate determinations in multicellular organisms (e.g., cell proliferation, differentiation, and apoptosis) are typical behaviors regulated by intracellular signal transduction processes \cite{Nishida1993,Saitoh1998,Butler1998,Chen2001,Chang2001,Qi2005}.
		These simultaneous internal biochemical processes require the synthesis and interaction of a number of different proteins on various biomembranes and in the cytoplasm involving several macromolecules, the cytoskeleton, and organelles.
		
		Recent studies suggest that the volume fraction of macromolecules in individual cells is much higher than that under typical {\it in vitro} conditions \cite{Fulton1982, Minton1992, Zimmerman1993, Minton1998, Minton2001, Ellis2001, Hall2003, Kim2009, Agrawal2009,Jiao2010a, Minton2010a,Fritsch2011, Fernandez2011a, Shtilerman2002,Nagarajan2011a,Cheung2005, Kinjo2002,Kilburn2010, Lin2009, Aoki2011a}.
		The total volume concentration of macromolecules in a typical cell is estimated to be 50--400 $\mathrm{mg/mL}$, whereas under typical {\it in vitro} conditions, it is estimated as 1--10$\mathrm{mg/mL}$ \cite{Fulton1982}.
		Such a high volume fraction of molecules, commonly called ``molecular crowding'', gives rise to extreme spatial restrictions.
		Thus, the diffusion and deformation (reaction) of molecules in the cytoplasm are highly suppressed \cite{Minton1992,Zimmerman1993, Minton1998,Minton2001, Ellis2001, Shtilerman2002, Hall2003,Kim2009, Agrawal2009,Jiao2010a, Minton2010a,Fritsch2011,Fernandez2011a, Nagarajan2011a}.
		Such spatial restrictions are also expected to enhance protein folding \cite{Cheung2005, Kinjo2002}, protein formation, stabilization of the intracellular architecture \cite{Kilburn2010}, and processive phosphorylation of ERK MAP kinase \cite{Lin2009, Aoki2011a}.
		
		The transduction of signals from the extracellular environment starts with the activation of receptors and signaling proteins on the cell membrane.
		Thus, the sensing and response of cells are dependent on the effective transport and reaction of signaling proteins in a 2-dimensional space.
		Recently, imaging measurements of macromolecules on the cell membrane have been performed extensively \cite{Lill2005, Daumas2003, Matsuoka2009, Miyanaga2007, Matsuoka2006, Saxton1997, Mashanov2006,Suzuki2005, Kusumi1993, Fujiwara2002, Nakada2003}.
		In some of these measurements, the typical motion of membrane proteins was revealed to be subdiffusive \cite{Kusumi1993, Suzuki2005, Fujiwara2002, Nakada2003}.
		This observation implies the existence of intrinsic membrane domains, such as raft or nonimaged molecules, which restrict the observed molecular motions by means of their excluded volumes.
		Thus, to better understand the performance and mechanisms of the upstream part of the signal transduction processes, the excluded volume effects of molecules on the reaction and diffusion dynamics of 2-dimensional systems should be clarified.
		
		In this paper, we investigated the influences of the excluded volume of molecules on biochemical processes using \textcolor{black}{a simple model} of typical signal transduction processes on the biomembrane.
		In the next section, we provide a brief introduction of G protein-coupled receptor (GPCR) signaling processes as a typical signal transduction process on the cell membrane and construct a model inspired by these processes.
		In the third section, we perform the simulation of the model and the analysis of a more simplified model.
		The summary and conclusion are presented in the final section.

	\section*{Model and Simulation Method}
		
		{\color{black}
		\subsection*{Typical signal transduction processes on the biomembrane}
		}
		In this section, we introduce a simple model of biochemical reaction processes that mimics the GPCR signaling processes on the biomembrane.
		GPCR signaling processes are typical signaling processes that play important roles in adaptations to environmental variations.
		The followings is a brief introduction to these processes.
		
		G proteins and GPCRs constitute large protein families of guanine nucleotide-binding proteins and their receptors \cite{Gilman1987,Wettschureck2005}.
		GPCRs sense extracellular signals (light-sensitive compounds, odors, pheromones, hormones, and neurotransmitters).
		The GPCRs activated by the extracellular signals activate G proteins by exchanging GTP in place of the GDP on G proteins.
		The activated G proteins usually separate into the $\alpha$ subunit ($G_\alpha$-$GTP$) and $\beta\gamma$ complex ($G_{\beta\gamma}$).
		Both $G_{\alpha}$-$GTP$ and $G_{\beta\gamma}$ activate different second messenger or effector proteins.
		The second messengers of some signaling pathways are located on the membrane and unbind from the membrane following activation in order to transfer the signals downstream of the signaling pathway to genes through lower hierarchical signal transductions.
		The GTP binding to $G_{\alpha}$ is hydrolyzed and becomes GDP, and then $G_{\alpha}$-$GDP$ binds to $G_{\beta\gamma}$.
		The inactivated G protein, {\it i.e.}, $G_{\alpha}$-$GDP$-binding $G_{\beta\gamma}$, can rebind to the GPCR.

		The abovementioned signaling processes can be summarized as the following:
		$i)$ The signaling proteins (G protein) activated by the receptor (GPCR) activate the target proteins (second messengers) on the membrane.
		$ii)$ The activated target proteins unbind from the membrane to transfer the signal downstream of the signaling pathways.
		A similar reaction cascade is also involved in the EGF-RAS-RAF signaling process on the membrane \cite{Dreux2006,Wong2004,Orton2005,Schulze2005}.
		Based on these facts, we constructed a simple model of the membrane signal transduction processes containing only typical molecular processes, as described in the next subsection.

		{\color{black}
		\subsection*{Reaction scheme of the model}
		}
		The model consisted of active and inactive receptors ($R^*$ and $R$), active and inactive signaling proteins ($S^*$ and $S$), bound and unbound target proteins ($T$ and $T^*$), and nonreactive molecules (crowder, $C$), which diffuse and react in a 2-dimensional space.
		Here, each molecule, $R$, $R^*$, $S^*$, $S$, $T$, and $C$, possessed its own volume.
		These molecules moved randomly under the restriction of their excluded volumes.
		Specifically, the distance between the centers of 2 molecules could not be smaller than the sum of their radii of inertia.
		
\begin{figure}[hpb]
			\begin{center}
				\includegraphics[keepaspectratio, width=2.7in]{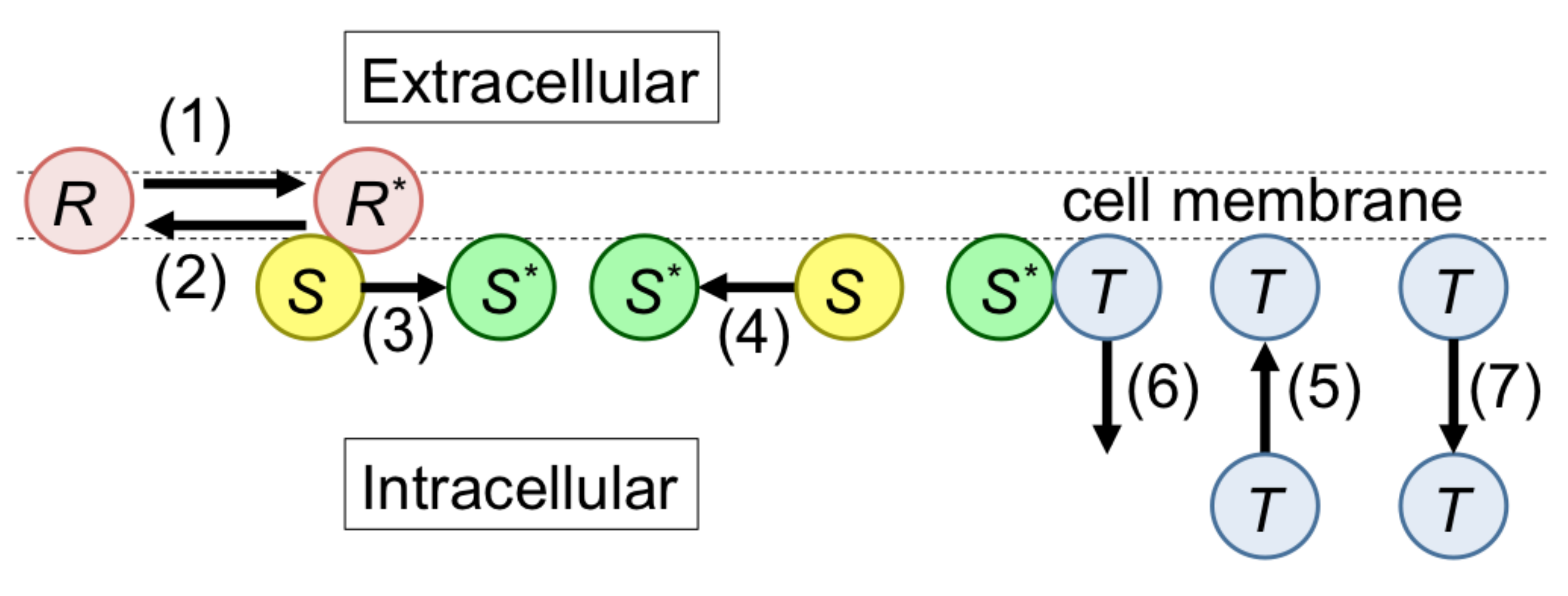}
				\caption{
					Illustration of the signaling pathway considered in this study.
					(1) The extracellular signal is transferred via activation of the receptor (Equation [Eq.] \ref{eq:activation_of_receptor}).
					(2) Autonomous inactivation of the receptor (Eq. \ref{eq:inactivation_of_receptor}).
					(3) Activation of the signaling protein by the active receptor (Eq. \ref{eq:activation_of_signaling}).
					(4) Autonomous inactivation of the signaling protein (Eq. \ref{eq:inactivation_of_signaling}).
					(5) Stochastic binding of the target protein from the cytoplasm to the membrane (Eq. \ref{eq:binding_of_target}).
					(6) Activation and unbinding of target proteins by active signaling proteins (Eq. \ref{eq:unbinding_of_target}).
					(7) Autonomous unbinding of the target protein (Eq. \ref{eq:unbinding_of_target2}).
				}
				\label{fig:signaling}
			\end{center}
		\end{figure}

		The signal transduction process is described as a cascade that follows the activation of receptors to the unbinding of activated target proteins from the membrane through the following reactions (Figure \ref{fig:signaling}).

		\noindent
		(A) A receptor autonomously changes from active to inactive, and vice versa, with the reaction rates $k_R$ and $k_{R^*}$, respectively, described by
		\begin{align}
			\label{eq:activation_of_receptor}
			R&\xrightarrow{k_R} R^*,
		\end{align}
		and
		\begin{align}
			\label{eq:inactivation_of_receptor}
			R^*&\xrightarrow{k_{R^*}} R.
		\end{align}
		\noindent
		(B) When an inactive signaling protein makes contact with an active receptor, this signaling protein is activated with the reaction rate $k_S$.
		The active signaling protein autonomously becomes inactive with the reaction rate $k_{S^*}$.
		These processes are described by
		\begin{align}
			\label{eq:activation_of_signaling}
			R^*+S&\xrightarrow{k_S} R^*+S^*,
		\end{align}
		and
		\begin{align}
			\label{eq:inactivation_of_signaling}
			S^*&\xrightarrow{k_{S^*}} S.
		\end{align}
		\noindent
		(C) If an empty space exists on the membrane, a target protein autonomously binds there with the \textcolor{black}{rate} $P_{in}$.
		When a target protein makes contact with an active signaling protein, this target protein is activated with reaction rate $k_T$.
		The target protein unbinds from the membrane as soon as it is activated.
		In the absence of activation, the target protein autonomously unbinds from the membrane with the reaction rate $P_{out}$.
		These processes are described by
		\begin{align}
			\label{eq:binding_of_target}
			\mbox{(empty space on membrane)}&\xrightarrow{P_{in}} T,
		\end{align}
		\begin{align}
			\label{eq:unbinding_of_target}
			&S^*+T\xrightarrow{k_{T}} S^*+T^*\nonumber\\
				&\to S^*~(T^*~\mbox{immediately unbinds from membrane}),
		\end{align}
		and
		\begin{align}
			\label{eq:unbinding_of_target2}
			T&\xrightarrow{P_{out}} (T~\mbox{unbinds from membrane}).
		\end{align}
		
		{\color{black}
			$P_{in}$ is proportional to the rate of target protein binding to lipids in the membrane when they collide with each other.
			This rate $P_{in}'$ in termed.
			$P_{in}'/P_{out}$ denotes the affinity between a target protein and the membrane.
			Moreover, $P_{in}$ is proportional to the number density and the diffusion rate of the target protein in the cytoplasm.
			Thus, $P_{in}$ indicates the effective binding rate of the target protein, which depends on the molecular species and cell species, and cell conditions.
			We assume that $P_{in}$ varies in the range of 0--1, where $P_{in}=1$ indicates that the $P_{in}'$ is large enough and/or that the number density of target proteins in the cytoplasm is high enough.
			We have also noted that the diffusions of molecules in the cytoplasm are much faster than those on the membrane.
			Subsequently, the target proteins in the cytoplasm tend to distribute uniformly and collide frequently with the membrane.
			Thus, in this model, we assumed that $P_{in}$ is uniform in space.
		}

	\subsection*{Cell-based model}
		
\begin{figure}[hpb]
			\begin{center}
				\includegraphics[keepaspectratio, width=1.5in]{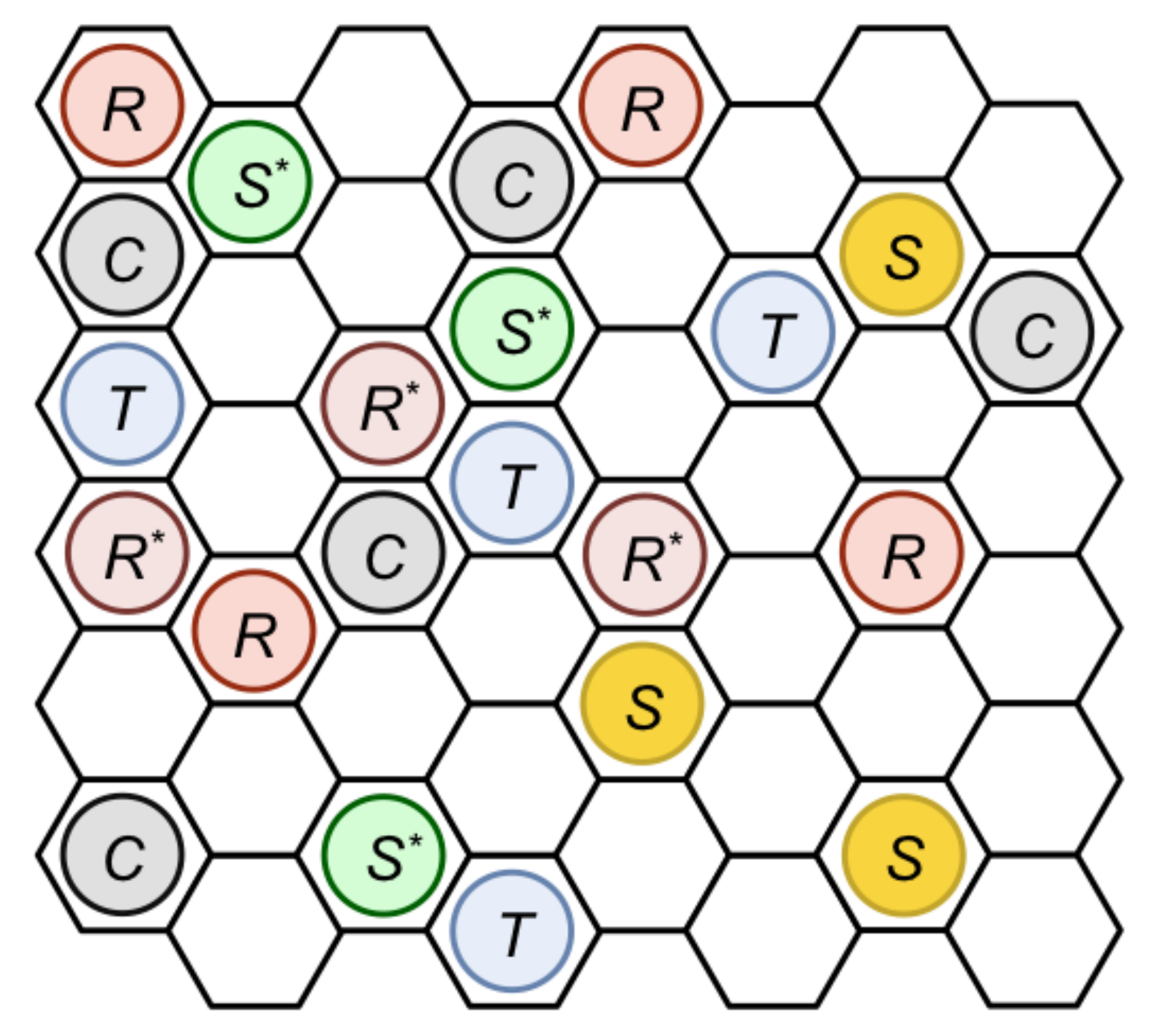}
				\caption{
					Illustration of the cell-based model.
					The membrane is described as a hexagonal lattice surface.
					Each cell can contain only one protein.
					$R$, inactive receptor; $R^*$, active receptor; $S$, inactive signaling protein; $S^*$, active signaling protein; $T$, target protein; and $C$, crowder.
				}
				\label{fig:model}
			\end{center}
		\end{figure}

		We have used a 2-dimensional cell-based model \cite{Takahashi2005,Bhide2000} to describe the diffusion and reactions of active and inactive receptors ($R^*$ and $R$), active and inactive signaling proteins ($S^*$ and $S$), target proteins ($T$), and crowders ($C$) on the membrane.
		The space was divided into $N$ 2-dimensional hexagonal cells as shown in Figure \ref{fig:model}.
		We defined the boundary condition of the system as periodic.
		Each cell could contain only one molecule, which represented the excluded volume effect.
		Each molecule randomly hopped from one cell to a neighboring empty cell or reacted in the manner indicated by Equations \ref{eq:activation_of_receptor}, \ref{eq:inactivation_of_receptor}, \ref{eq:inactivation_of_signaling}, \ref{eq:binding_of_target}, or \ref{eq:unbinding_of_target2}.
		Reactions \ref{eq:activation_of_receptor}, \ref{eq:inactivation_of_receptor}, \ref{eq:inactivation_of_signaling}, and \ref{eq:unbinding_of_target2} occurred spontaneously with the given reaction rates, whereas 2-body reactions \ref{eq:activation_of_signaling} and \ref{eq:unbinding_of_target} occurred when 2 corresponding molecules existd in adjacent cells.
		No reaction occurred by the crowders.
		In the empty cells, reaction \ref{eq:binding_of_target} occurred at the rate $P_{in}$.
		
	\subsection*{Simulation method}
		
		To simulate the present cell-based model, we used the Monte-Carlo method. 
		The temporal evolution of the system progressed by the iteration of the following steps.
		\\
		$(0)$ $R$ and $S$ are distributed randomly to yield the initial condition.
		\\
		$(i)$ One of the cells is chosen randomly.
		\\
		$(ii)$ If this cell contains molecule $S$ or $T$, the corresponding 2-body reactions, \ref{eq:activation_of_signaling} or \ref{eq:unbinding_of_target}, occur at a rate determined by the product of its given [reaction rates] $\times$ [the number density of the corresponding catalyst on the six neighboring cells].
		$T^*$ is removed from this cell as soon as it appears.
		\\
		$(iii)$ If this cell contains molecule $R$, $R^*$, $S^*$, or $T$, the corresponding reaction, \ref{eq:activation_of_receptor}, \ref{eq:inactivation_of_receptor}, \ref{eq:inactivation_of_signaling}, or \ref{eq:unbinding_of_target2}, occurs with its respective reaction rate.
		Reaction \ref{eq:unbinding_of_target2} indicates that $T$ is removed from this cell.
		\\
		$(iv)$ If this cell contains a molecule but no reaction occurs, this molecule moves randomly to one of the 6 neighboring cells as long as the chosen cell is empty.
		\\
		$(v)$ If this cell contains no molecule, the binding process of a target protein \ref{eq:binding_of_target} occurs with the reaction rate $P_{in}$.
		This indicates that $T$ becomes bound to this cell with the reaction rate $P_{in}$.
		
		In each time step, $(i)$--$(v)$ are iterated $N$ times, where $N$ is the number of cells.
		\textcolor{black}{We defined the time step of the system as $t$ when $(i)$--$(v)$ were iterated $tN$ times from the initial condition.}
		In this study, we assumed the length of each cell was $\sim 1 \mathrm{nm}$, and order of unit time step was $\sim 10\mbox{--}100 \mathrm{\mu s}$ (details are provided in Appendix B).
		
		We definde the signal flow, $J$, as the average frequency of target proteins per receptor.
		$J$ at time $t$ is derived from [Number of activations of target proteins between $t$ and $t$ + 1]/[Number of receptors].
		We also defined the ``occupancy'' of molecules $\rho$ as [Total number of molecules in the system]/$N$, and the occupancy of molecule $X$ as $[X]=\mbox{[Number of molecule}~X~\mbox{in the system]}/N$.
		In the cell-based model, this value was used as an index of the crowding of molecules instead of the volume fraction of molecules frequently measured in experiments.
		It was assumed that the volume fraction of molecules in the system and their occupancy in the corresponding cell-based model were positively correlated.
		The rough estimations of the volume fractions of molecules from the occupancy in simple molecular systems are stated in Appendix B.
		
		\textcolor{black}{
			Recently, there have been few experimental observations of the total volume fractions of molecules on cell membranes.
			However, such aspects are naturally expected to depend on the specificity of molecules around the membrane and on cellular conditions.
			In the present model, the effects of such specificities are described by the parameters $P_{in}$ and $P_{out}$, and the occupancy of signaling proteins remaining on the membrane.
			Thus, in the present study, we systematically varied these parameters in order to consider the possible reaction behaviors on a biomembrane in several possible situations.
		}

	\section*{Results and Discussion}
	
		\subsection*{Simulation result}
		
			In this section, we consider the typical properties obtained through the simulation of the model, which did not include crowders.
			We focus on the steady-state signal flow $J$, defined as the average frequency of the target proteins activations per receptor, and the total occupancy of molecules, $\rho$, for several values of signaling protein occupancy, $[S_{tot}] = [S] + [S^*]$, the effective binding rate of the target protein, $P_{in}$, and autonomous unbinding rate of the target protein, $P_{out}$.
			For simplicity, some parameters were fixed: $k_R=1$, $k_{R^*}=0$, $k_S = 0.3$, $k_{S^*} = 0.3$, $k_T = 0.3$ and $N=1600$.
			\textcolor{black}{
			Here, $k_R=1$ and $k_{R^*}=0$ indicate that all receptors are always activated.
			The qualitative results are unaffected by these parameters if $k_R$ is large enough and $k_{R^*}$ is small enough, {\it i.e.} signals are input frequently from outside the cell.
			}
			We also assumed that the occupancy of the receptor was a low value, $[R_{tot}]=0.01$, as recent experimental observations have reported that the volume fraction of receptors on the cell membranes is estimated as a few percent \cite{Janssens1987,VanderWoning2009,Shankaran2006,Mayawala2005,Ozcan2006,Macdonald2008}.
			However, the following arguments are qualitatively independent of these details.
			
			\textcolor{black}{
			If we assume that the effect of the spatial distribution and excluded volume of molecules can be neglected, the signal flow is \textcolor{black}{obtained} by the mean-field analysis as
			\begin{align}
				\label{eq:mean_main}
				J &= k_T[T][S^*] 
				= \frac{a P_{in} (1 - [R_{tot}] - [S_{tot}])}{P_{in} + P_{out} + a},\nonumber\\
				a&=\frac{k_Rk_S k_T[R_{tot}] [S_{tot}]}{k_Rk_{S^*}+k_{R^*}k_{S^*}+k_Rk_S[R_{tot}]}.
			\end{align}
			Here, the derivation of this form is provided in Appendix A.
			This result indicates that $J$ is a monotonically increasing function of $P_{in}$ that takes place in the form of a Michaelis-Menten-type equation independent of the values of the reaction rates, $[S_{tot}]$, and $P_{out}$.
			On the other hand, the simulations results of the presented model deviate considerably from those expected by mean-field analysis, as described below.
			}
			
\begin{figure}[hpb]
			\begin{center}
				\includegraphics[keepaspectratio, width=2.3in]{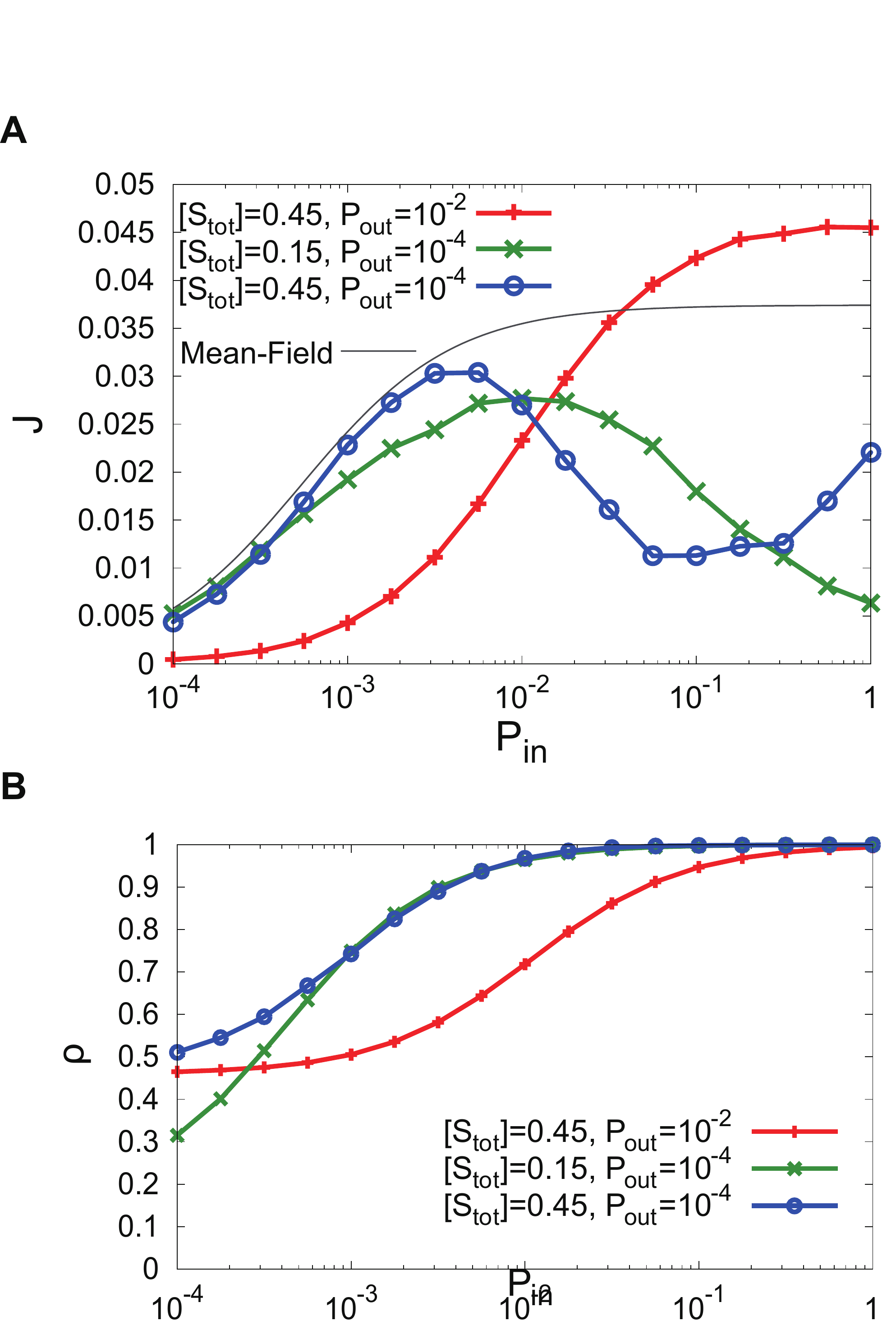}
				\caption{
					\textcolor{black}{(A) Signal flow $J$ and (B) total occupancy $\rho$ as functions of the binding rate of the target protein $P_{in}$ for ($[S_{tot}], P_{out})=(0.45, 10^{-2})$ (red plus), $(0.15,10^{-4})$ (green cross), $(0.45, 10^{-4})$ (blue circle) obtained by simulations, and that for ($[S_{tot}], P_{out})=(0.45, 10^{-2})$ obtained by the mean field analysis (black line).}
				}
				\label{fig:result_lattice_gas}
			\end{center}
		\end{figure}

\begin{figure}[hpb]
			\begin{center}
				\includegraphics[keepaspectratio, width=2.2in]{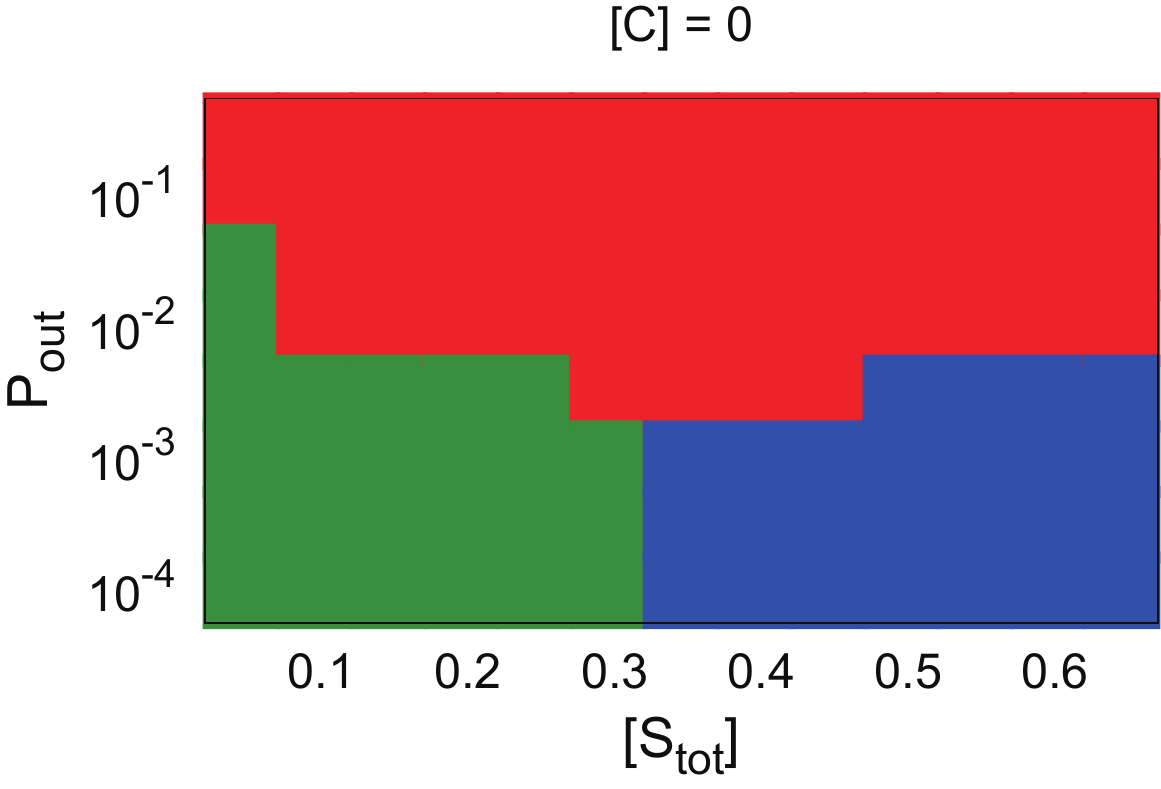}
				\caption{
					Phase diagram of $J$--$P_{in}$ relation for each $[S_{tot}]$ and $P_{out}$ obtained by simulation.
					Red: $J$ monotonically increases, Green: $J$ exhibits a bell-shaped curve, Blue: $J$ exhibits an S-shape curve with an increase in $P_{in}$.
				}
				\label{fig:result_lattice_gas_phase}
			\end{center}
		\end{figure}

			Figure \ref{fig:result_lattice_gas}A and \ref{fig:result_lattice_gas}B depict $J$ and $\rho$ as functions of $P_{in}$ obtained by the simulation for the parameter sets $([S_{tot}], P_{out})=(0.45, 10^{-2})$ (red plus), $(0.15, 10^{-4})$ (green cross), and $(0.45, 10^{-4})$ (blue circle).
			As shown in Figure \ref{fig:result_lattice_gas}, unlike the result obtained by the mean-field analysis, there are 3 typical $J$ variations with the increase in $P_{in}$,
			$i)$ increasing monotonically,
			$ii)$ increasing then decreasing in a bell-shaped curve, and
			$iii)$ increasing, decreasing, then increasing in an S-shaped curve.
			Figure \ref{fig:result_lattice_gas_phase} illustrates the phase diagram of the $J$--$P_{in}$ relationship at each $[S_{tot}]$ and $P_{out}$.
			Here, $J$ exhibits a monotonic increase for a case of large $P_{out}$,
			a bell-shaped curve when both the $[S_{tot}]$ and $P_{out}$ are small, and
			an S-shaped curve for the case of a large $[S_{tot}]$ and small $P_{out}$.

		\subsection*{Spatial organization}
			
			The finding in the previous subsection implies the existence of a spatially nonuniform distribution of molecular species.
			Thus, to observe the characteristic spatial distributions of molecules, we measured the radial distribution function of each molecular species around each receptor.
			The radial distribution functions of the signaling proteins and target proteins around the receptor, $d_S(r)$, and $d_T(r)$, are defined as
			\begin{align}
				\label{eq:radial_distribution_function_S}
				d_S(r)&=\frac{\langle\rho_S(r)+\rho_{S^*}(r)\rangle}{[S_{tot}]},\\
				\label{eq:radial_distribution_function_T}
				d_T(r)&=\frac{\langle\rho_T(r)\rangle}{\langle [T]\rangle	},
			\end{align}
			respectively.
			Here, $\rho_X(r)$ $(X = \{S, S^*, T\})$ indicates the local occupancy of molecule $X$ at a distance $r$ from the receptor (see Appendix C).
			$\langle~\rangle$ Donates the sample and long time-averaged value.
			Molecule $X$ is considered dense when $d_X(r) >1$, and sparse when $d_X(r) <1$, compared to the uniform distribution.
			
\begin{figure}[hpb]
			\begin{center}
				\includegraphics[keepaspectratio, width=2.7in]{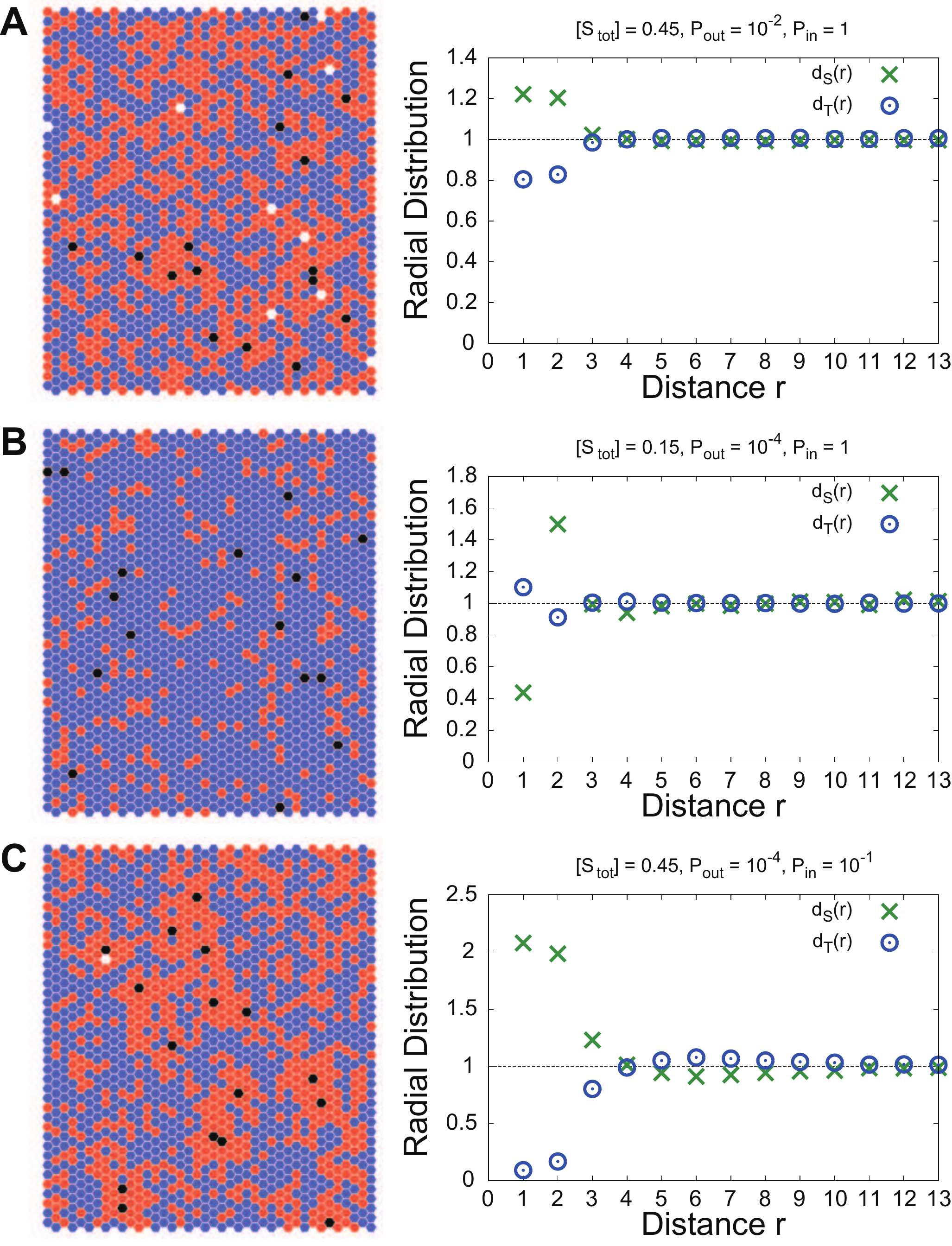}
				\caption{
					(Left) Typical snapshots of the present simulation, (Right) radial distribution of signaling proteins (green cross) and target proteins (blue circle) as a function of the distance from a receptor for $([S_{tot}], P_{in}, P_{out}) = $(A) $(0.45, 1, 10^{-2})$, (B) $(0.15, 1, 10^{-4})$, (C) $(0.45, 10^{-1}, 10^{-4})$.
					Each black, orange, and blue point indicates a receptor, signaling protein, and target protein, respectively.
				}
				\label{fig:radial_structure}
			\end{center}
		\end{figure}

			Figure \ref{fig:radial_structure} depicts typical snapshots of the simulation and radial distributions of the signaling protein $d_S(r)$ (green cross), and the target protein $d_T(r)$ (blue circle) for the following cases:
			(A) $([S_{tot}], P_{out}, P_{in})=(0.45, 10^{-2}, 1)$ at which $J$ realizes the largest value in the case that $J$ monotonically increases with $P_{in}$,
			(B) $([S_{tot}], P_{out}, P_{in}) = (0.15, 10^{-4}, 1)$ at which $J$ decreases along the bell-shaped curve,
			and (C) $([S_{tot}], P_{out}, P_{in})=(0.45, 10^{-4}, 10^{-1})$ at which $J$ yields a local minimum of the S-shaped curve.
			It should be noted that snapshots and radial distributions similar to Figure \ref{fig:radial_structure}A can be obtained for parameter sets in which $J$ is at the peak of the bell-shaped or S-shaped curve.
			In these cases, the molecules tend to distribute according to the following spatial structure:
			the signaling proteins surround the receptor to form the receptor-signaling protein cluster ($R$-$S$ cluster), and the target proteins become distributed around such clusters.
			If the molecular distribution occurs according to the abovementioned structure, $S$ around $R^*$ and $T$ around $S^*$ tends to be activated rapidly.
			Subsequently, the reaction process of the system progresses actively.
			
			{\color{black}
			The qualitative mechanism of the formation of $R$-$S$ clusters is explained as follows.
			Around $R$, $S$ is frequently activated to $S^*$, and $T$ near $S$ is also frequently activated.
			The activated $T$ unbinds from the membrane.
			Then, empty spaces appear around $S^*$, {\it i.e.}, near $R$.
			These empty spaces are occupied by other molecules according to the diffusion or binding of $T$.
			Through these processes, the molecular flow in which molecules approach $R$ tends to be formed.
			At the terminal of this molecular flow, $T$ tends to unbind by the activation, but $S$ ($S^*$) remains near $R$.
			Thus, $R$-$S$ clusters are formed.
			
			In the case of appropriate values of $P_{in}$ and $P_{out}$, in which $[T]$ is not as large as compared to $[S_{tot}]$, $T$ is distributed around the $R$-$S$ clusters.
			However, other types of molecular distribution often appear, in particular, in the case of a small $P_{out}$.
			For example, in the case of a large $P_{in}$ and small $[S_{tot}]$, in which $[S_{tot}]$ is much smaller than $[T]$, $R$ tends to be surrounded not by $S$ but $T$, as in Figure \ref{fig:radial_structure}B.
			On the other hand, in the case of a large $[S_{tot}]$, the $R$-$S$ cluster tends to be surrounded by $S$, as in Figure \ref{fig:radial_structure}C.
			
			Now, we explain the behaviors of the present model qualitatively, based on the abovementioned molecular distributions.
			First, we considered the case of a small $P_{out}$ and small $[S_{tot}]$.
			If the $P_{in}$ is too small, where the total occupancy of molecules is so small that the excluded volume effects can be neglected, $J$ increases with $P_{in}$, as considered in the mean-field analysis.
			Moreover, in the case of a not-so-large $P_{in}$, $[T]$ appears an appropriate values compared to $[S_{tot}]$;
			$J$ increases with $P_{in}$ because $R$-$S$ clusters appear and the $T$ surrounding them increases.
			However, with the increase in $P_{in}$, $[T]$ becomes so much larger than $[S_{tot}]$ that $T$ tends to surround between $R$ instead of $S$.
			Thus, $J$ increases then decreases in a bell-shaped curve with the increase in $P_{in}$.
			On the other hand, if the $P_{out}$ is large enough, an $R$-$S$ cluster can be formed even in the case of a large $P_{in}$ because $T$ around $R$ often unbinds and $S$ can approach $R$ to surround them.
			
			Next, we considered the case of a small $P_{out}$ and large $[S_{tot}]$.
			Similar to the above case, if the $P_{in}$ is small enough, $J$ increases with $P_{in}$.
			On the other hand, with the increase in the total occupancy of molecules by the increase in $P_{in}$, $R$-$S$ clusters surrounded by $S$ appear.
			Here, $S$ around the $R$-$S$ clusters is usually inactive because it cannot be activated by $R$.
			If $P_{in}$ is not so large that the total occupancy of molecules is not large either, $S$ can surround the $R$-$S$ cluster and $T$ can exchange their positions by their diffusion.
			Then, $T$ can be activated and $J$ increases with $P_{in}$.
			On the other hand, such diffusions and exchanges tend to be suppressed with the increase in $P_{in}$.
			Then, $J$ decreases with the increase in $P_{in}$ in a range of not-small-enough $P_{in}$.
			However, for a much larger $P_{in}$, $T$ can invade the void between the $R$-$S$ cluster and $S$ around the cluster by binding from the cytoplasm and being activated as soon as such a void appears.
			Here, such voids are created by the fluctuation of the reaction and diffusion of molecules on the membrane.
			If the $P_{out}$ is large enough, $S$ around the $R$-$S$ cluster and $T$ can exchange positions smoothly even in the case of a large $P_{in}$ because $T$ near $S$ around the $R$-$S$ cluster often unbinds, and $S$ can diffuse.
			Then, $T$ often approach $R$-$S$ clusters and $J$ always increases with $P_{in}$.
			
			The above qualitative considerations are consistent with the mathematical analysis of the present model through the more simplified model, as mentioned in the next subsection.
			In the present argument, we assume that the signaling protein cannot unbind from the membrane.
			However, the qualitative results are unchanged even when the signaling proteins can autonomously unbind from the membrane if the unbinding rate of the signaling proteins is small enough compared to that of the target proteins.
			}

		\subsection*{Theoretical analysis by a simple stochastic model}
			The results of the previous subsection indicate that the structure of the molecular distribution around the receptor forms a dominant contribution to the reaction activity of the present system.
			In this subsection, we analyze a simple 1-dimensional stochastic model that describes the molecular motions in the radial direction around a receptor.
			The analysis is then compared to 3 described $J$--$P_{in}$ relationships.
			\textcolor{black}{
			For simplicity, we consider the case that only one receptor exists in the system \textcolor{black}{and is always activated, {\it i.e.} $R^*$}.
			}
			
\begin{figure}[hpb]
			\begin{center}
				\includegraphics[keepaspectratio, width=2.7in]{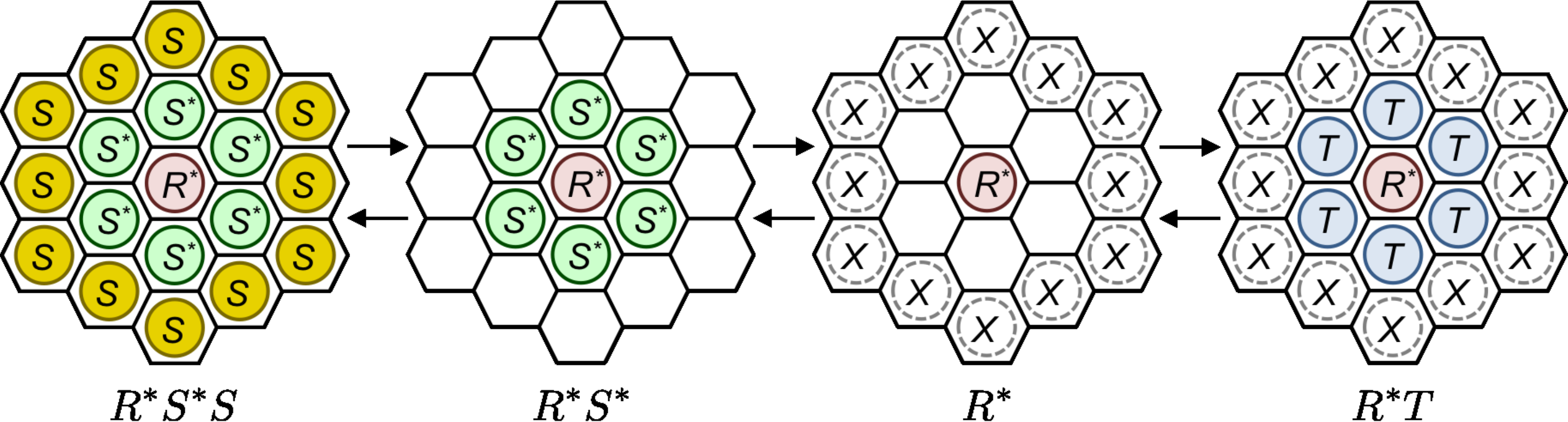}
				\caption{
					{\color{black}
					Considered states of the simple stochastic model and corresponding molecular distributions images in 2-dimensional space.
					$R^*$, the receptor; $S$, inactive signaling protein; $S^*$, active signaling protein; and $T$, target protein; each $X$ denotes $R^*$, $S$, $S^*$, $T$, or empty.}
				}
				\label{fig:illustration_spatial_ordering_analysis}
			\end{center}
		\end{figure}

			We consider the following 3 typical states of the molecular distributions in the radial direction around a receptor as:
			``$R^*$'', where no molecule exists beside the receptor; ``$R^*T$'', where the target protein exists beside the receptor; and the state where the signaling protein exists beside the receptor and is activated.
			The third state is divided into the following 2 states:
			``$R^*S^*$'', where no molecule exists beside the active signaling protein,
			and ``$R^*S^*S$'', where an inactive signaling protein exists beside the active signaling protein.
			Here, in the states ``$R^*$'', ``$R^*T$'', and ``$R^*S^*S$'', any reactions can not occur.
			On the other hand, in the ``$R^*S^*$'' state, the reaction occurs if a target protein appears beside $S^*$.
			{\color{black}
			Note that other states exist, such as ``$R^*S^*T$'', ``$R^*S^*S^*$'' and ``$R^*S$''.
			To simplify, however, we have omitted these states from this analysis by considering the following assumptions.
			$i)$ The target protein beside $S^*$ is rapidly activated and unbound.
			$ii)$ The signaling protein becomes inactive immediately if it leaves the receptor.
			$iii)$ The signaling protein is activated as soon as it approaches next to the receptor.
			
			Moreover, we assumed that the molecular distributions around a receptor are almost uniform in terms of the angle direction as shown in Figure \ref{fig:illustration_spatial_ordering_analysis}, which is roughly supported by the simulation results in the previous subsections.
			Then, we only considered the molecule reactions and diffusions only in the radial direction.
			Although these assumptions render the model too simple, the appearance of the 3 types of $J$--$P_{in}$ relationships obtained in the previous subsection can be explained qualitatively as follows.
			
			}

			Now, we consider the following transition dynamics among these 4 states,
			\begin{align}
				R^*S^*S\overset{k_1}{\underset{k_{-1}}{\rightleftharpoons}}R^*S^*\overset{k_2}{\underset{k_{-2}}{\rightleftharpoons}} R^* \overset{k_3}{\underset{k_{-3}}{\rightleftharpoons}} R^*T.
				\label{eq:default}
			\end{align}
			where $k_1$, $k_{-1}$, $k_2$, $k_{-2}$, $k_3$, and $k_{-3}$ indicate the transition rates.
			These transition rates are approximately estimated by $[S_{tot}]$, $[T]$, $P_{in}$, and $P_{out}$ as follows.
			
			$k_1$: The transition from ``$R^*S^*S$'' to ``$R^*S^*$'' indicates that $S$ diffuses away from beside $S^*$.
			The transition rate is proportional to the probability that any molecules do not exist beside $S$.
			Then,
			\begin{align}
				k_1=(1-[S_{tot}]-[T])p_d,
				\label{eq:state_k1}
			\end{align}
			where $p_d(<1)$ indicates the diffusivity of molecules.
			
			$k_{-1}$: The transition from ``$R^*S^*$'' to ``$R^*S^*S$'' indicates that $S$ diffuses from an adjacent space to occupy the empty space beside $S^*$.
			This transition rate is proportional to the probability that no $T$ appears, but $S$ does.
			The probability of the appearance of $S$ is estimated as $[S_{tot}]p_d$, and that of $T$ is estimated as $P_{in}+(1-P_{in})[T]p_d$, which is the sum of contributions by the binding from the cytoplasm and the diffusion of target proteins on the membrane.
			Then, 
			\begin{align}
				k_{-1}&=p_d[S_{tot}]\{1-[P_{in}+(1-P_{in})p_d[T]]\}\nonumber\\
				&=p_d[S_{tot}](1-p_d[T])(1-P_{in}).
				\label{eq:state_k-1}
			\end{align}
			
			$k_2$: The transition from ``$R^*S^*$'' to ``$R^*$'' indicates that $S^*$ leaves from beside $R^*$ before any molecules occupy the empty space beside $S^*$.
			The probability that 1 molecule ($S$ or $T$) appears at the empty space beside $S^*$ is estimated by $P_{in}+(1-P_{in})([S_{tot}]+[T])p_d$.
			Then,
			\begin{align}
				k_2&=\{1-[P_{in}+(1-P_{in})([S_{tot}]+[T])p_d]\}p_d\nonumber\\
				&=p_d(1-p_d[S_{tot}]-p_d[T])(1-P_{in}).
				\label{eq:state_k2}
			\end{align}
			
			$k_{-2}$: The transition from ``$R^*$'' to ``$R^*S^*$'' indicates that $S$ diffuses from an adjacent space to occupy the empty space beside $R^*$.
			The properties of this transition are almost the same as those of ``$R^*S^*$'' to ``$R^*S^*S$''.
			Then,
			\begin{align}
				k_{-2}&=p_d[S_{tot}]\{1-[P_{in}+(1-P_{in})p_d[T]]\}\nonumber\\
				&=p_d[S_{tot}](1-p_d[T])(1-P_{in}).
				\label{eq:state_k-2}
			\end{align}
			
			$k_3$: The transition from ``$R^*$'' to ``$R^*T$'' indicates that $T$ originates from the cytoplasm or adjacent spaces on the membrane to occupy the empty space beside $R^*$.
			This transition rate is proportional to the probability that no $S$ appears but $T$ does.
			Then,
			\begin{align}
				k_{3}&=(1-p_d[S_{tot}])[P_{in}+(1-P_{in})p_d[T]].
				\label{eq:k3}
			\end{align}
			
			$k_{-3}$: The transition from ``$R^*T$'' to ``$R^*$'' indicates that $T$ leaves from beside $R^*$.
			This transition rate is yielded by the sum of 2 contributions: the autonomous unbinding from the membrane and the diffusion of $T$.
			The diffusion of $T$ is proportional to the probability that any molecules do not exist beside $T$.
			Then,
			\begin{align}
				k_{-3}&=P_{out}+(1-P_{out})(1-[S_{tot}]-[T])p_d.
				\label{eq:state_k-3}
			\end{align}
			
			Here, it should be noted that while $[S_{tot}]$ and $P_{in}$ and $P_{out}$ are the control parameters of the present system,
			$[T]$ should be derived from these parameters.
			In the following analysis, with reference to the simulation results in Figure \ref{fig:result_lattice_gas}B, we assume $[T]$ is derived as
			\begin{align}
				{[T]}=\frac{(1-[S_{tot}])P_{in}}{\alpha [R_{tot}] [S_{tot}] +P_{out}+P_{in}},
				\label{eq:state_target}
			\end{align}
			with the fitting parameter $\alpha\sim1/6$.
			
			The steady-state probability distribution of the 4 considered states, $Q_{R^*S^*S}, Q_{R^*S^*}, Q_{R}, Q_{R^*T}$ are obtained by
			\begin{align}
				Q_{R^*S^*S} = \frac{k_{-1} k_{-2} k_{-3}}{k_{-1} k_{-2} k_{-3}+k_1 k_{-2} k_{-3}+k_1 k_2 k_{-3}+k_1 k_2 k_3},\\
				Q_{R^*S^*} = \frac{k_1 k_{-2} k_{-3}}{k_{-1} k_{-2} k_{-3}+k_1 k_{-2} k_{-3}+k_1 k_2 k_{-3}+k_1 k_2 k_3},\\
				Q_{R^*} = \frac{k_1 k_2 k_{-3}}{k_{-1} k_{-2} k_{-3}+k_1 k_{-2} k_{-3}+k_1 k_2 k_{-3}+k_1 k_2 k_3},\\
				Q_{R^*T} = \frac{k_1 k_2 k_3}{k_{-1} k_{-2} k_{-3}+k_1 k_{-2} k_{-3}+k_1 k_2 k_{-3}+k_1 k_2 k_3}.
				\label{eq:state_steady_solution}
			\end{align}
			The steady-state signal flow is estimated by $Q_{R^*S^*}$ $\times$ [the probability of appearance of the target protein beside $S^*$].
			The latter probability is that for which no $S$ appears but $T$ does, by diffusion or binding from the cytoplasm.
			Then, $J=(1-p_d[S_{tot}])\{P_{in}+(1-P_{in}) p_d[T]\}Q_{R^*S^*}=k_3Q_{R^*S^*}$.
					
\begin{figure}[hpb]
			\begin{center}
				\includegraphics[keepaspectratio, width=2.7in]{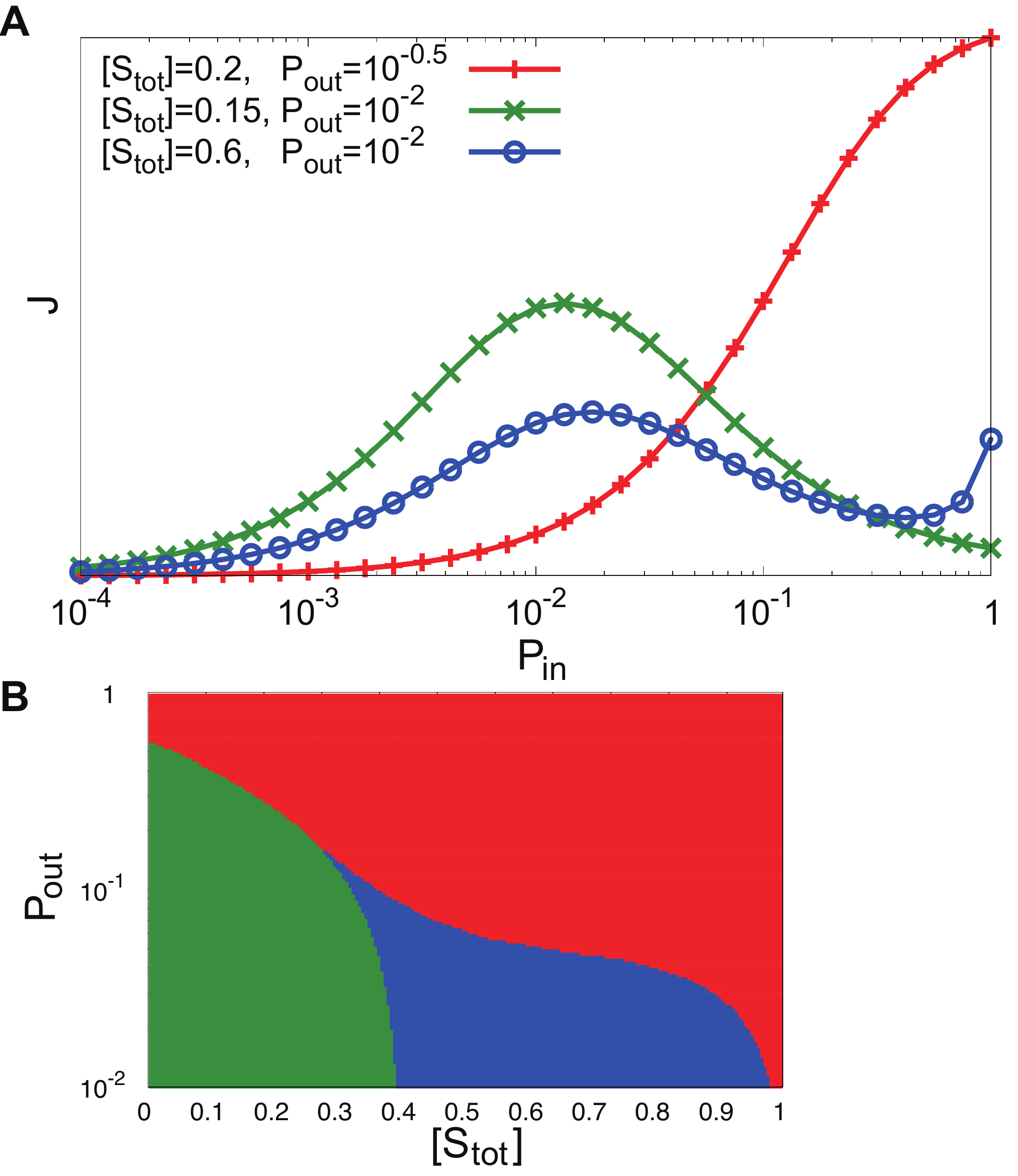}
				\caption{
					(A) Signal flow $J$ as a function of $P_{in}$ for ($[S_{tot}], P_{out})=(0.2, 10^{-\frac{1}{2}})$ (red plus), $(0.15,10^{-2})$ (green cross), and $(0.6, 10^{-2})$ (blue circle);
					(B) phase diagram of $J$--$P_{in}$ relation for each $[S_{tot}]$ and $P_{in}$ obtained by analysis of the stochastic model.
					Each symbol is plotted in the same manner as Figure \ref{fig:result_lattice_gas_phase}.
				}
				\label{fig:result_spatial_ordering_analysis}
			\end{center}
		\end{figure}

\begin{figure}[hpb]
			\begin{center}
				\includegraphics[keepaspectratio, width=3.6in]{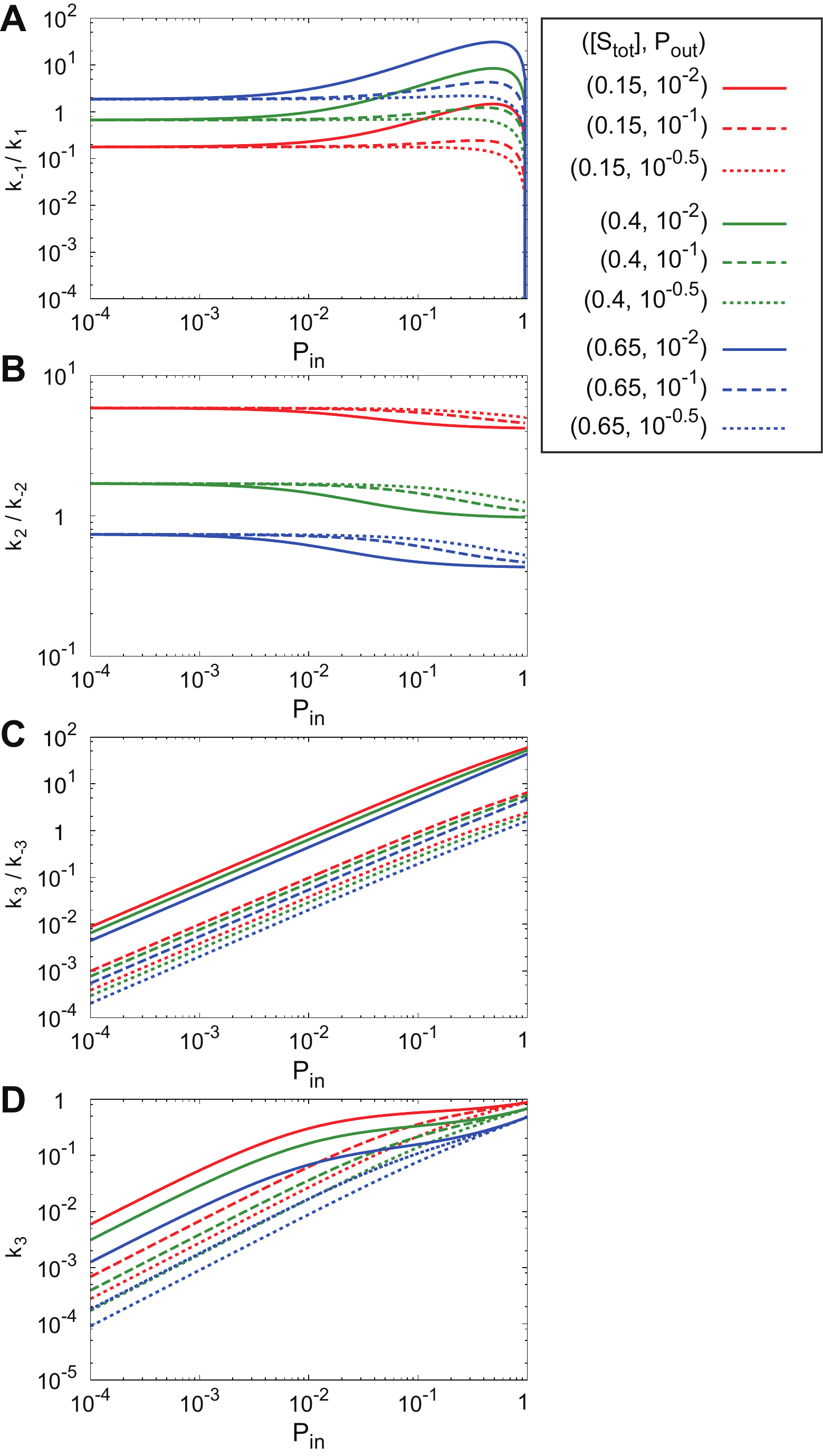}
				\caption{
					{\color{black}(A) $k_{-1}/k_1$, (B) $k_{2}/k_{-2}$, (C) $k_{3}/k_{-3}$, and (D) $k_3$ as functions of $P_{in}$ for combinations of $[S_{tot}]=\{0.15$(red), $0.4$ (green), $0.65$ (blue)$\}$ and $P_{out}=\{10^{-2}$ (solid line), $10^{-1}$ (dashed line), $10^{-\frac{1}{2}}$ (dotted line)$\}$.}
				}
				\label{fig:pin_dependency_of_rates}
			\end{center}
		\end{figure}

			Figure \ref{fig:result_spatial_ordering_analysis} depicts (A) $J$ as a function of $P_{in}$ for some parameter sets of $([S_{tot}], P_{out})$, and (B) the phase diagram of $J$--$P_{in}$ relationships against $[S_{tot}]$, and $P_{out}$ for $p_d=0.8$, obtained by analyzing the steady state solutions of the present stochastic model.
			These results are qualitatively independent of $p_d$.
			As shown in these figures, we obtained results that were qualitatively similar to the simulation results in the previous subsections.
			{\color{black}
			
			Then, we considered the detailed properties of this model.
			It should be noted that $J$ is also described as
			\begin{align}
				J=\frac{k_3}{1+\cfrac{k_{-1}}{k_1}+\cfrac{k_{2}}{k_{-2}}+\cfrac{k_{2}}{k_{-2}}\cfrac{k_{3}}{k_{-3}}},
				\label{eq:signal_flow}
			\end{align}
			which indicates that $J$ depends on the ratios between the transition rates, $k_{-1}/k_1$, $k_{2}/k_{-2}$, $k_{3}/k_{-3}$, and $k_3$.
			Then, the appearances of the 3 types of $J$-$P_{in}$ dependency can be explained by considering the $P_{in}$ dependencies of $k_{-1}/k_1$, $k_{2}/k_{-2}$, $k_{3}/k_{-3}$ and $k_3$ as follows.
			
			Figure \ref{fig:pin_dependency_of_rates} illustrates (A) $k_{-1}/k_1$, (B) $k_{2}/k_{-2}$, (C) $k_{3}/k_{-3}$ and (D) $k_{3}$ as functions of $P_{in}$ for combinations of $[S_{tot}]=\{0.15 \mbox{(red)}, 0.4 \mbox{(green)}, 0.6 \mbox{(blue)}\}$ and $P_{out}=\{10^{-2} \mbox{(solid line)}, 10^{-1} \mbox{(dashed line)}, 10^{-\frac{1}{2}} \mbox{(dotted line)}\}$, where their analytic forms are obtained by
			
			\begin{align}
				\label{eq:k-1_k1}
				\frac{k_{-1}}{k_1}=\frac{[S_{tot}](1-p_d[T])(1-P_{in})}{1-[S_{tot}]-[T]},
			\end{align}
			\begin{align}
				\label{eq:k2_k-2}
				\frac{k_2}{k_{-2}}&=\frac{p_d(1-p_d[S_{tot}]-p_d[T])(1-P_{in})}{p_d[S_{tot}](1-p_d[T])(1-P_{in})}\nonumber\\
				&=\frac{1}{[S_{tot}]}-\frac{p_d}{1-p_d[T]},
			\end{align}
			\begin{align}
				\label{eq:k3_k-3}
				\frac{k_{3}}{k_{-3}} = \frac{(1-p_d[S_{tot}])\{P_{in}+(1-P_{in})p_d[T]\}}{P_{out}+(1-P_{out})(1-[S_{tot}]-[T])p_d},
			\end{align}
			and Equation (\ref{eq:k3}).
			Figure \ref{fig:pin_dependency_of_rates}B and \ref{fig:pin_dependency_of_rates}C and Equations (\ref{eq:k2_k-2}) and (\ref{eq:k3_k-3}) indicate that
			$k_2/k_{-2}$ is a monotonically decreasing function of $P_{in}$ and a monotonically increasing function of $P_{out}$,
			and that $k_3/k_{-3}$ is almost proportional to $P_{in}$ and a monotonically decreasing function of $P_{out}$.
			This indicates that, with the increase in $P_{in}$, $R^*$ tends to make contact with other molecules because the occupancy, {\it i.e.}, the volume fraction, of molecules on the membrane increases.
			In the same manner, $k_3$ increases monotonically with $P_{in}$ as indicated in Figure \ref{fig:pin_dependency_of_rates}D and Equation (\ref{eq:k3}).
			However, the slope of each curve varies from steep to gradual because $[T]$ increases with $P_{in}$ but is saturated for a large $P_{in}$.
			
			On the other hand, as shown in Figure \ref{fig:pin_dependency_of_rates}A, the variations of $k_{-1}/k_1$ are somewhat more complicated, as follows.
			Equation (\ref{eq:k-1_k1}) indicates that $k_{-1}/k_1$ has a maximum at
			\begin{align}
				P_{in}=\frac{1 - \alpha [R_{tot}]}{2} -\frac{ P_{out}}{2 [S_{tot}]}
			\end{align}
			for a large $[S_{tot}]$ and small $P_{out}$ that $P_{out}<[S_{tot}](1-\alpha [R_{tot}])$ satisfies.
			On the other hand, with an increase in $P_{in}$, $k_{-1}/k_1$ decreases monotonically for $P_{out}>[S_{tot}](1-\alpha [R_{tot}])$.
			Furthermore, it increases with the increase in $P_{out}$.
			The appearance of this maximum of $k_{-1}/k_1$ for a small $P_{out}$ means that with the increase in $P_{in}$, the ``$R^*S^*S$'' state tends to occur for a small or intermediate value of $P_{in}$, but the transition rate to the ``$R^*S^*S$'' state tends to be hindered for a large $P_{in}$.
			The reason for this is considered to be as follows.
			With the increase in the $P_{in}$, the occupancy of molecules becomes large.
			Then, the molecules tend to make more contact with each other, inducing the transition from ``$R^*S^*$'' to ``$R^*S^*S$''.
			On the other hand, if $P_{in}$ becomes much larger, $T$ can approach beside $R^*S^*$ more frequently (and is activated and unbinds immediately) by the binding from the cytoplasm before $S$ diffuses close to $R^*S^*$.
			With the increase in $P_{in}$, $k_{-1}/k_1$ then increases for a not-so-large $P_{in}$, but decreases for a large $P_{in}$.
			
			According to the abovementioned $P_{in}$ dependencies of each term, the appearances of the 3 $J$-$P_{in}$ relationsships are explained as follows:
			
			$i)$ For a large $P_{out}$, $k_{-1}/k_1$ and $k_{2}/k_{-2}$ are monotonically decreasing functions of $P_{in}$.
			Even if we assume that $k_{-1}/k_1$ and $k_{2}/k_{-2}$ are constant values, $J$ is given as a monotonically increasing function of $P_{in}$ in the form of a Michaelis-Menten-type equation.
			According to these facts, $J$ increases monotonically with $P_{in}$.
			
			$ii)$ For a small $P_{out}$ and small $[S_{tot}]$, $k_{2}/k_{-2}$ is sufficiently larger than $k_{-1}/k_1$ and $k_{3}/k_{-3}$ for a small $P_{in}$.
			Then, only $k_{2}/k_{-2}$ is the dominant term of the denominator of Equation (\ref{eq:signal_flow}).
			In this case, the denominator of Equation (\ref{eq:signal_flow}) is considered constant.
			Then, $J$ increases with $P_{in}$ for a small $P_{in}$, since $k_3$ monotonically increases with $P_{in}$.
			On the other hand, $k_{3}/k_{-3}$ increases with $P_{in}$, and finally exceeds $1$.
			Then, $(k_{2}/k_{-2}) \times (k_3/k_{-3})$ also becomes the dominant term of the denominator of Equation (\ref{eq:signal_flow}).
			Here, the slope of $k_3$ becomes gradual with the increase in $P_{in}$, while the slope of $k_3/k_{-3}$ is unchanged and $k_{2}/k_{-2}$ is regarded as constant for a large $P_{in}$.
			Therefore, $J$ decreases with the increase in $P_{in}$ for a large $P_{in}$.
			$J$ exhibits the bell-shaped curve.
			
			$iii)$ For a small $P_{out}$ but large $[S_{tot}]$, $k_{-1}/k_1$ becomes dominant as compared to $k_2/k_{-2}$ and $k_3/k_{-3}$ for a small $P_{in}$.
			Here, $k_{-1}/k_1$ is approximately constant for a small $P_{in}$.
			Then, the dominator of Equation (\ref{eq:signal_flow}) is considered constant.
			On the other hand, $(k_{2}/k_{-2}) \times (k_3/k_{-3})$ increases to the same order as that of $k_{-1}/k_1$ with the increase in $P_{in}$.
			Then, $(k_{2}/k_{-2}) \times (k_3/k_{-3})$ also becomes the dominant term of the denominator of Equation (\ref{eq:signal_flow}).
			Thus, in a similar manner to case $ii)$, $J$ increases and turns to decrease with $P_{in}$.
			However, if $P_{in}$ approaches 1, $k_{-1}/k_1$ exhibits a drastic decrease where the slope of the decrease in $k_{-1}/k_1$ is much steeper than that of the increase in $(k_{2}/k_{-2}) \times (k_3/k_{-3})$.
			Then, the denominator turns to decrease in the neighborhood of $P_{in}=1$, and $J$ increases again with $P_{in}$.

			Furthermore, the $P_{in}$ at which $J$ exhibits a maximum peak is obtained by the intersection of $k_2/k_{-2}$ and $(k_2/k_{-2})\times (k_3/k_{-3})$, {\it i.e.}, $\sim P_{out}+\alpha [R_{tot}][S_{tot}]$.
			Thus, the $P_{in}$ that exhibits the peak shifts to the larger value with the increase in $P_{out}$.
			On the other hand, the $P_{in}$ at which $J$ exhibits the local minimum of the S-shaped curve slightly decreases with the increase in $P_{out}$.
			Here, the local minimum value of $J$ increases with the increase in $P_{out}$.
			If $P_{out}$ increases more, the local minimum of $J$ vanishes as shown in Figure \ref{fig:result_spatial_ordering_phase_boundary}.

			}

		\subsection*{Reaction system with crowding molecules}
			
			In the previous subsections, we considered a simple model of an ideal situation that was assumed to comprise only the components of the signal transduction processes.
			However, in general, several reaction processes take place simultaneously on the cell membrane, using several macromolecules.
			The components of these other reactions often behave as obstacles for the components of other reaction processes.
			Thus, to elucidate the influences of the excluded volume of molecules using a more realistic model, we simulated a system containing a crowder molecule, $C$.
			Here, $C$ moved randomly on the membrane without reaction, binding, or unbinding.
			It only hindered the random movements of other reactive molecules because of its excluded volume.
			
\begin{figure}[hpb]
			\begin{center}
				\includegraphics[keepaspectratio, width=3.7in]{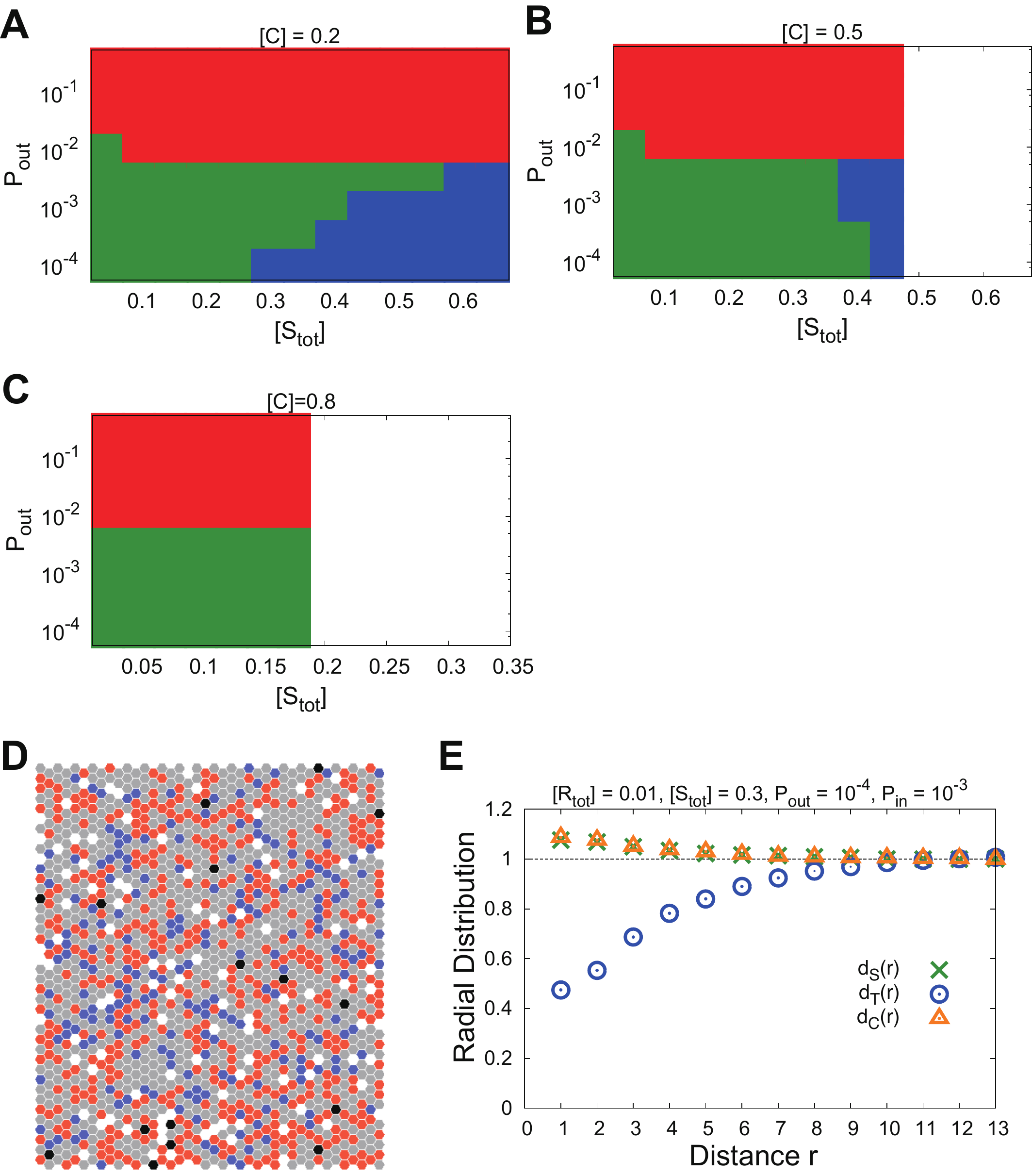}
				\caption{
					Phase diagram of $J$--$P_{in}$ relation for each $[S_{tot}]$ and $P_{out}$ obtained by simulations
					(A) $[C]=0.2$, (B) $[C]=0.5$, and (C) $[C]=0.8$.
					(D) Typical snapshot of the simulation, (E) radial distributions of signaling proteins (green cross), target proteins (blue circle), and crowder (yellow triangle) plotted as a function of the distance from a receptor for $P_{in}= 10^{-3}$, $P_{out}=10^{-4}$, $[S_{tot}]= 0.3$, and $[C]=0.5$, which are plotted in the same manner as in Figure \ref{fig:radial_structure}.
					Each black, orange, blue, and gray point indicates a receptor, signaling protein, target protein, and crowder, respectively.
				}
				\label{fig:result_lattice_gas_crowding2D_1}
			\end{center}
		\end{figure}

			Figure \ref{fig:result_lattice_gas_crowding2D_1} depict the $J$-$P_{in}$ relation for (A) $[C]=0.2$, (B) $[C] = 0.5$ and (C) $[C] = 0.8$ with $[R_{tot}] = 0.01$, described in the same manner as Figure \ref{fig:result_lattice_gas_phase}, \textcolor{black}{(D) a typical snapshot of the simulation, and (E) the radial distributions of the signaling proteins $d_S(r)$, the target proteins $d_T(r)$, and the crowder $d_C(r)$ for $P_{in}= 10^{-3}$, $P_{out}=10^{-4}$, $[S_{tot}]= 0.3$, and $[C]=0.5$, described in the same manner as Figure \ref{fig:radial_structure}}.
			As shown in Figure \ref{fig:result_lattice_gas_crowding2D_1}A and \ref{fig:result_lattice_gas_crowding2D_1}B, the phase diagrams for not too many large values of $[C]$ are qualitatively the same as those obtained when $[C]=0$ (no crowders).
			\textcolor{black}{We also observed that the signaling proteins tended to distribute around the receptor on average, as shown in Figure \ref{fig:result_lattice_gas_crowding2D_1}D and \ref{fig:result_lattice_gas_crowding2D_1}E, when the system exhibited a large $J$, similar to the $R$-$S$ cluster formation observed in the case of $[C]=0$.}
			
			{\color{black}
				However, in the phase diagrams, the phase boundary between the regions with bell-shaped and S-shaped $J$--$P_{in}$ relations shifted to a large $[S_{tot}]$ with an increase in $[C]$.
				Moreover, in cases of a much larger $[C]$, such as $[C]=0.8$, the region with the S-shaped relation disappeared, as shown in Figure \ref{fig:result_lattice_gas_crowding2D_1}C.
				The reason for these facts is believed to be as follows.
				The appearance of the S-shaped relation was caused by the aggregation of $S$ to form the $R$-$S$ cluster surrounded by $S$, as mentioned in the previous subsections.
				However, with the increase in $[C]$, such aggregations of $S$ tended to be hindered and required more $S$ to be formed.
				Thus, the phase boundary between the regions with bell-shaped and S-shaped $J$-$P_{in}$ relations shifted to a large $[S_{tot}]$.
				Moreover, if $[C]$ became much larger, $[S_{tot}]$ could not be so large as to form such aggregations of $S$.
				Hence, the region with the S-shaped relations disappeared.
			}

	\section*{Summary and Conclusion}

		We investigated the influences of the excluded volume of molecules on the activity of reaction processes on 2-dimensional surfaces using a cell-based model of signal transduction processes on biomembranes.
		The simulation was based on the diffusion and reaction among receptors, signaling proteins, target proteins, and crowders on a 2-dimensional surface.
		
		{\color{black}
		With the increase in the binding frequency of target proteins to the membrane, the volume fraction of molecules on the membrane increased in a similar manner to the molecular crowding in the cytoplasm.
		However, the reaction behaviors on such a 2-dimensional membrane were obtained differently from those in a 3-dimensional bulk system.
		}
		We found that the signal flow exhibited 3 types of molecular volume fraction dependencies according to the abundance ratio and the binding/unbinding rate of the molecules constructing the system:
		$i)$ When the autonomous unbinding of target proteins occurred frequently, the signal flow increased monotonically with the binding rate of the target proteins.
		$ii)$ When the autonomous unbinding of target proteins occurred rarely and the number of signaling proteins was small, the signal flow increased for the small values of the binding rate, and then decreased for the large value of the binding rate of the target proteins.
		$iii)$ When the autonomous unbinding of target proteins occurred rarely and the number of signaling proteins was sufficiently large, the signal flow increased and decreased as in $(ii)$, but then increased again for a sufficiently large value of the binding rate of the target proteins.
		We further demonstrated that the excluded volume of molecules influenced their hierarchical distributions throughout the reaction processes.
		In particular, when the system exhibited a large signal flow, the signaling proteins tended to surround the receptors and the target proteins tended to distribute around the receptor--signaling protein clusters, which accelerated the activations of the signaling proteins and target proteins.
		
		To control the signal transduction activity on the membrane, we expect that the formation of the presented hierarchical molecular distributions makes a dominant contribution along with receptor clustering \cite{Bray1998}.
		On the other hand, a large number of reaction processes other than the signaling cascade are known to take place on the interior and exterior surfaces of several biomembranes of organelles, such as the mitochondria, Golgi body, and nucleus.
		{\color{black}
		The molecular distributions on the majority of such membranes are not clearly understood experimentally, except the recent reports of the aggregations of peptides related to some diseases \cite{Ashkenazi1998a,Shtilerman2002,White2010a}.
		Such studies are progressing as ongoing issues or will be studied as future issues.
		For such problems,} based on our present argument and extended arguments with more realistic models, we predict that several molecule aggregation patterns are formed by the excluded volume of molecules and influence on the function of reaction networks on several biomembranes.

	\section*{Acknowledgements}
		The authors are grateful to Y. Togashi, K. Takahashi, K. Aoki, M. Nishikawa and M. Kikuchi for the fruitful discussion and useful information.

	\appendix
	\section*{Appendix}
	\subsection*{A. Mean-field approximation}
		\renewcommand{\theequation}{A\arabic{equation}}
		\setcounter{equation}{0} 
		
		We analyzed the present model by mean-field approximation based on the mass action law.
		For simplicity, we showed only the results of the system without crowders because the results were qualitatively the same as that for the system with crowders.
		The temporal evolution of the occupancy of each molecular species, $[R]$, $[R^*]$, $[S^*]$, $[S ]$, and $[T]$, is obtained by
		\begin{align}
			\label{eq:ri}
			\frac{d [R] }{dt}&=-k_R[R]+k_{R^*}[R^*]\\
			\label{eq:ra}
			\frac{d [R^*] }{dt}&=k_R[R]-k_{R^*}[R^*]\\
			\label{eq:si}
			\frac{d [S] }{dt}&=-k_S[R^*][S]+k_{S^*}[S^*]\\
			\label{eq:sa}
			\frac{d [S^*] }{dt}&=k_S[R^*][S]-k_{S^*}[S^*]\\
			\label{eq:ti}
			\frac{d[T]}{dt}&=(1-\rho)P_{in}-k_T[T][S^*]-P_{out}[T]\\
			\label{eq:rho}
			\rho&= [S_{tot}]+[R_{tot}]+[T].
		\end{align}
		
		From these equations, we estimated the steady-state signal flow, defined as the frequency of activation for unbinding of the target protein, given by
		\begin{align}
			\label{eq:mean}
			J &= k_T[T][S^*] 
			= \frac{a P_{in} (1 - [R_{tot}] - [S_{tot}])}
					{P_{in} + P_{out} + a},\nonumber\\
			a&=\frac{k_Rk_S k_T[R_{tot}] [S_{tot}]}{k_Rk_{S^*}+k_{R^*}k_{S^*}+k_Rk_S[R_{tot}]}.
		\end{align}
		This equation clearly exhibits a Michaelis-Menten-type equation in terms of $P_{in}$, subsequently, $J$ monotonically increases with $P_{in}$ independent of $[S_{tot}]$ and $P_{out}$.
		
	\subsection*{B. Spatial-temporal scale and volume fraction of the present model}
		\renewcommand{\theequation}{B\arabic{equation}}
		\setcounter{equation}{0} 
		
		We considered the relationship between the spatial-temporal scales of the present model and those of experiments.
		First, we considered the unit length and the unit time in the present model.
		
		If we regard the present model as the GPCR signal transduction system, its spatial scales are estimated by considering the size of each cell to be of the same order as its molecular size, i.e., $\mathrm{nm}$.
		The diffusion rate of molecules was estimated as $\sim 10^{-1}$--$10^{-2} \mathrm{\upmu m^2/s} = 10^{-1}$--$10^{-2} \mathrm{nm^2/\upmu s}$ for the GPCR without the cytoskeleton \cite{Suzuki2005}.
		Hence, the time interval in which a molecule moves a single molecule length is estimated as $\sim 10$--$100 \mathrm{\upmu s}$.
		This time interval is the unit time step in the model.
		We also assumed that the $0.01$--$1$ reactions occur for each molecule at each time step.
		This assumption yields the characteristic time for reactions as $0.1$--$10 \mathrm{ms}$, which is consistent with the time scale of the conformational change of typical proteins.
		
		Next, we considered	 the relationship between occupancy in the present argument and the volume fraction of molecules.
		In general, the detailed relationship between the two is nontrivial because it depends on the detailed properties of the considered system, such as the sizes and shapes of the molecules and environmental components.
		However, if all molecules are considered spheres with almost the same radii, the volume fraction is roughly estimated from the occupancy, as in the following.
		
		In the present model, each cell could contain only one molecule.
		This implies that the size of each cell was larger than that of each molecule, but had to be small enough for the distance between 2 molecules in neighboring cells to be always smaller than the molecular diameter, to avoid the possibility of the invasion of a molecule between 2 neighboring molecules.
		Then, the length of the diagonal of each cell, $d$, and the radius of the molecules, $r$, always had to satisfy the equation
		\begin{align}
			\sqrt{\left(\frac{\sqrt{3}d}{2}-r\right)^2 + \left(\frac{d}{4}-\frac{r}{\sqrt{3}}\right)^2}-r \le r.
			\label{eq:spatial_scale1}
		\end{align}
		According to such $d$ and $r$, the ratio of the volume fraction to the occupancy is obtained by
		\begin{align}
			\frac{\mbox{[Volume fraction]}}{\mbox{[Occupancy]}}=\frac{\mbox{[area of one molecule]}}{\mbox{[area of one cell]}}=\frac{8\pi r^2}{3\sqrt{3}d^2}.
			\label{eq:spatial_scale2}
		\end{align}
		
		For example, we assumed that the length of the diagonal of each hexagonal cell was $\sim 1.7 \times$ [molecular diameter], which is close to the largest value to satisfy Equation \ref{eq:spatial_scale1}.
		Then, the volume fraction of the densest condition with occupancy = 1 in the 2-dimensional cell-based model was estimated as $\sim 41$\%, which is a slightly large but possible value in experimental situations \cite{Fulton1982}.
		
	\subsection*{C. Details of quantification}
		In the simulation, the local occupancy of molecule $X$, $\rho_X(r)$, was calculated as follows.
		First, we defined the distances from the receptor to the molecules of the cell as the minimum number of steps to move from the receptor to the molecules.
		Second, we used $N_X(r)$ and $N(r)$ as the number of molecule $X$ and cells at a distance $r$ from the receptor to the molecules, respetively.
		Then,
		\begin{align}
			\rho_S(r) = \frac{N_X(r)}{N(r)}
		\end{align}

	\bibliography{M-Fujii}

		

		\begin{figure}[hpb]
			\begin{center}
				\includegraphics[keepaspectratio, height=4.0in, width=2.7in]{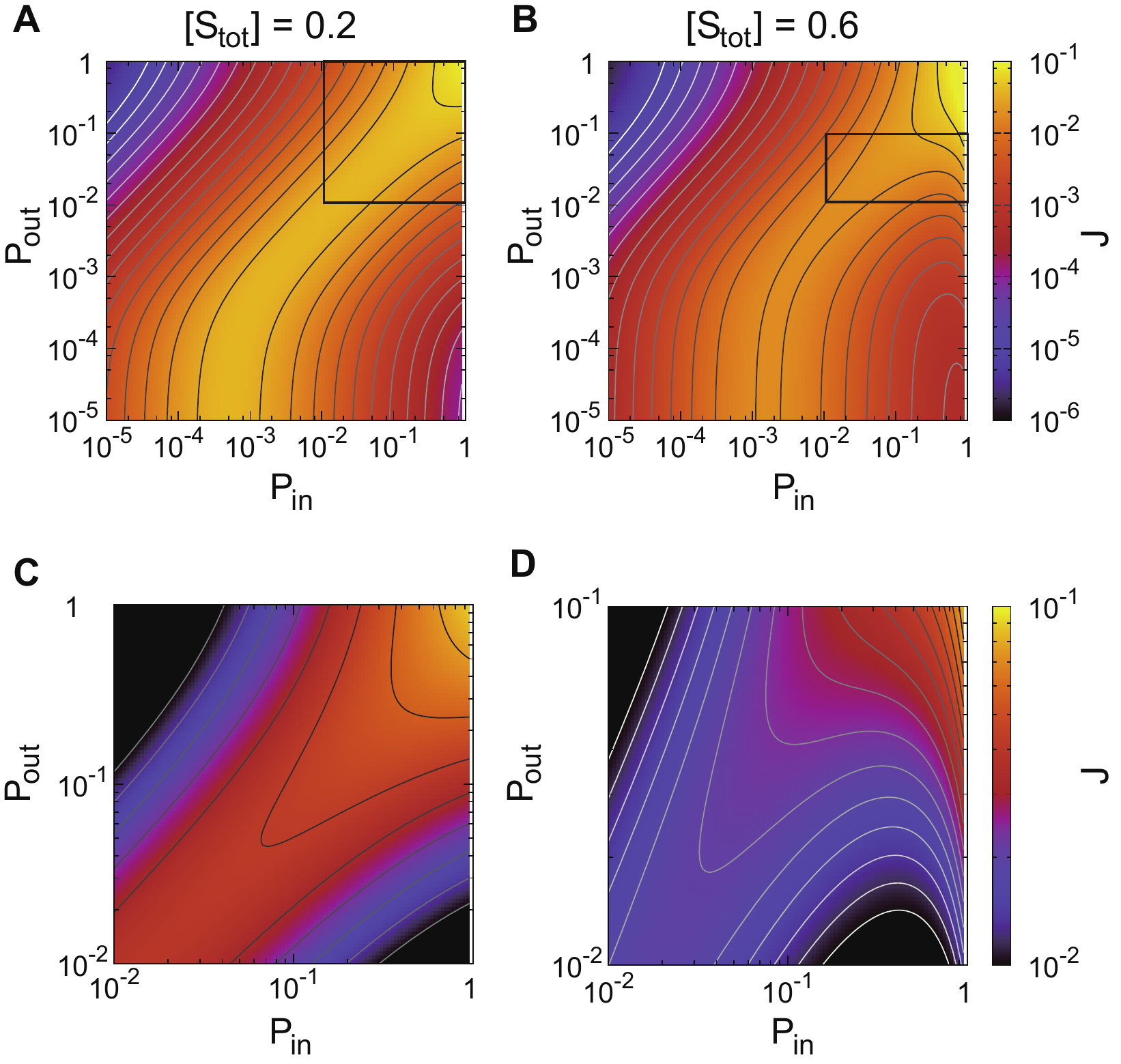}
				\caption{
					{\color{black}Signal flow $J$ as functions of $P_{in}$ and $P_{out}$ obtained by analysis of stochastic model.
					(A) $[S_{tot}]=0.2$, (B) $[S_{tot}]=0.6$.
					(C) and (D) are the enlarged insets in (A) and (B), respectively.}
				}
				\label{fig:result_spatial_ordering_phase_boundary}
			\end{center}
		\end{figure}
		
\end{document}